\documentclass[twocolumn,showpacs,preprintnumbers,amsmath,amssymb,pra]{revtex4}
\usepackage{txfonts}
\usepackage{graphicx,hyperref}
\usepackage{dcolumn}
\usepackage{bm}

 \allowdisplaybreaks
\begin{document}


\title{Analytical results for the superflow of spin-orbit-coupled  Bose-Einstein condensates in optical lattices}
\author{Xiaobing Luo$^{1,2,3}$}
\altaffiliation{Corresponding author: xiaobingluo2013@aliyun.com}
\author{Zhou Hu$^{1}$}
\author{Zhao-Yun Zeng$^{2}$}
\author{Yunrong Luo$^{4}$}
\author{Baiyuan Yang$^{2}$}
\author{Jinpeng Xiao$^{2}$}
\author{Lei Li$^{2}$}
\author{Ai-Xi Chen$^{1}$}

\affiliation{$^{1}$Department of Physics, Zhejiang Sci-Tech University, Hangzhou, 310018, China}
\affiliation{$^{2}$School of Mathematics and Physics, Jinggangshan University, Ji'an 343009, China}
\affiliation{$^{3}$State Key Laboratory of Low-Dimensional Quantum Physics, Department of Physics, Tsinghua University, Beijing 100084, China}
\affiliation{$^{4}$ Department of Physics and Key Laboratory for Matter Microstructure and Function of Hunan Province, and Key Laboratory of
Low-dimensional Quantum Structures and Quantum Control of Ministry
of Education, Hunan Normal University, Changsha 410081, China}

\date{\today}

\begin{abstract}
In this paper, we show that for
sufficiently strong atomic interactions, there exist analytical solutions of current-carrying
nonlinear Bloch states at the Brillouin zone edge to the model of spin-orbit-coupled  Bose-Einstein condensates (BECs) with symmetric spin interaction loaded into optical lattices. These simple but generic exact solutions provide an analytical demonstration of some intriguing properties which have neither an analog in the
regular BEC lattice systems nor in the uniform spin-orbit-coupled BEC systems. It is an analytical example for understanding the superfluid and other related properties of the spin-orbit-coupled BEC lattice systems.
\end{abstract}

\maketitle
\section{introduction}
As an intrinsic interaction between the motion and spin of a
particle, spin-orbit (SO) coupling in solid-state materials
plays an important role in many interesting physical phenomena and applications, such as topological
insulators\cite{Qi}, spin Hall effect\cite{Kato}, and spin devices\cite{Zutic}, etc. Recently, artificial SO-coupling has been successfully realized in both bosonic and fermionic
ultracold atomic systems\cite{Lin}-\cite{Wu2016}, which provides a brand new platform to explore
these exotic states and SO-coupled superfluids. The SO-coupled atomic systems
 exhibit many novel phases that have no parallel in conventional condensed
matter physics. For a Bose-Einstein condensate (BEC) in homogeneous space,
the experimentally accessible spin-orbit coupling  gives rise
to a class of nontrivial superfluid phase\cite{CWang}-\cite{CJWu} and collective excitations\cite{Khamehchi,SCJi}.
For example, the ground-state
phase diagram for a homogeneous spin-orbit-coupled BEC is demonstrated to include a stripe phase with periodic density modulation
that breaks continuous translation symmetry\cite{CWang,Ho,YLi}.

SO-coupled BECs in optical lattices have also received considerable attentions\cite{Cai}-\cite{LZhou}, due to the rich emergent physics
stemming from the interplay between SO
coupling and lattice effects. In a recent experiment, SO-coupled BECs in an optical lattice have been successfully
loaded into the destined nonlinear Bloch band, and the dynamical stabilities of Bloch states crucial to the
superfluidity of a BEC have been measured\cite{Hamner}, which provides a direct observation
of the broken Galilean invariance. On the theoretical side, the flat band dispersion of such SO-coupled lattice systems  has been uncovered\cite{YZhang2013} and a
full ground state phase diagram in the superfluid regime has been analyzed by numerical studies\cite{ZChen}.

For the fundamental
importance it is extremely valuable to have analytic
solutions to the theoretical model of BECs in optical lattices, in that they can provide a
deeper understanding of the underlying physics than straight
numerical simulations. The research activities of finding exact analytic solutions to regular BECs in periodic potentials
have been under way for a long time and a large family of exact
stationary solutions have been derived. Several known models that have exact analytical solutions include the one-component BEC in quasi-one-dimensional Kronig-Penney potential\cite{Seaman,WDLi}, and quasi-one-dimensional (or two-dimensional) Jacobi elliptic potential (which can be reduced to a sinusoidal potential in the limit case)\cite{Bronski1}-\cite{Deconinck}. These analytic results have been  generalized to multi-component (including two-component) regular BECs without SO coupling\cite{Kostov,Hai1}. Due to the advantage of being analytically tractable, these accurate solutions can serve
as good examples for understanding the appearance of  loops (swallowtails) in
the Bloch band structure which are related to superfluid
properties of the BEC\cite{Wu1}-\cite{Koller}. For example, Wu and Niu have used a simple exact nonlinear Bloch solution (which connects
the loop to the rest of the Bloch band)
to illustrate the multiple-valuedness of lowest band at the
Brillouin zone edge which makes it different
from the linear Bloch band\cite{Wu1,Wu2}.

Interestingly, such a looplike dispersion relation has also been found recently in a
SO-coupled BEC in free space\cite{YongpingZhang2019}. However, the presence of SO coupling poses a challenge to
finding the exact solutions to the BEC systems with periodic potentials and the
corresponding exact solutions to such systems remain extremely rare.
Recently, a type of  spatiotemporal Bloch state of  two-component SO-coupled BEC
in a high-frequency driven optical lattice has been obtained analytically, which nevertheless needs fine tuning of the system parameters to
match the rigorous equilibrium condition and therefore has no generic nature\cite{Hai2}.
Thus, it is highly desirable and worthwhile to seek for  generic exact solutions to the systems of  SO-coupled
BECs in optical lattices, which will certainly help to better understand
the superfluid and other physical properties
of SO-coupled lattice BEC systems.

In this paper, we present a set of analytical degenerate Bloch wave solutions for SO-coupled
BEC in an optical lattice.  These analytic solutions  exist only
for sufficiently large interaction strength and connect to a known
solution in the limit of zero SO coupling. Their properties related to superfluidity are fully analyzed.
 It is analytically demonstrated that SO coupling adds some results to the superfluidity of condensates at the
Brillouin zone edge. For example, these exact nonlinear Bloch wave states can carry a pure spin current and no
total atomic density current.
They also provide an analytical evidence for the looplike
Bloch band structure in the SO-coupled BEC lattice system.
By using the standard Bogoliubov
theory, these exact solutions are examined and shown to have both Landau and dynamical stabilities in certain parameter regions.

\section{Model equation}
We consider the mean-field model of a quasi-one-dimensional
SO-coupled BEC in the presence of an optical lattice along $x$ direction given by a
periodic potential $V(x)=V_0\sin^2(k_L x)$, where $k_L=2\pi/\lambda_L$ is the wavenumber of the laser beam with $\lambda_L$ being the laser wavelength, and $V_0$ is the lattice
depth. The dynamics of such a SO-coupled BEC system
can be described by the dimensionless Gross-Pitaevskii (GP) equation,
\begin{align}\label{GP}
i\frac{\partial  \Psi}{\partial t}  =[\hat{H} _{\rm{soc}}+V_0\sin^2(x)+\hat{H} _{\rm{non}} ]\Psi,
\end{align}
where $\Psi= (\Psi_1,\Psi_2)^T$ is the spinor describing the two pseudospin components of the BEC, $\hat{H} _{\rm{soc}}$ is
the single-particle SO-coupled Hamiltonian,
\begin{align}\label{Hsoc}
\hat{H} _{\rm{soc}}=& \frac{\hat{p}_x^{2} }{2}+k_{0}\hat{p}_x\hat{\sigma} _{z}+\frac{\Omega }{2}\hat{\sigma}_x,
\end{align}
 and the nonlinear terms originating from the atomic
interactions are explicitly given by
\begin{align}\label{Hnon}
\hat{H} _{\rm{non}}=& \begin{pmatrix}
 g_{11}\left | \Psi _{1}  \right | ^{2}+g_{12} \left | \Psi _{2}  \right | ^{2}&&0  \\0&
  &g_{22}\left | \Psi _{2}  \right | ^{2}+g_{12} \left | \Psi _{1}  \right |^{2}
\end{pmatrix}.
\end{align}
The model \eqref{GP} has been realized in the very recent experiment by Hamner \emph{et al}\cite{Hamner}.
In the above dimensionless GP equation, $\hat{p}_x=-i\partial_x$ is the atomic momentum operator, $\hat{\sigma}_{x,y,z}$ are the usual $2\times2$ Pauli matrices, $\Omega$ is the Raman frequency, and $k_0=k_{R}/k_L$ characterizes the
SO coupling strength which is determined by the wave number of the Raman
laser $k_R$. The units of energy, time, and length are chosen as $2E_L=\hbar^2k_L^2/m$ with $m$ being the atomic mass,
$m/(\hbar k_L^2)$, and $1/k_L$, respectively. The effective one-dimensional
atomic interactions is given by $g_{ij}=\hbar\omega_{\bot}n_0a_{ij}/E_L$ ($i,j=1,2$),
where $a_{ij}$ is the s-wave scattering length of spin components $i$ and $j$,
$\omega_{\bot}$ is the
the transversal trapping frequency, and $n_0=N/L_x$ is the averaged BEC density with $N$ being the atom number in one unit
cell and $L_x=\lambda_L/2=\pi/k_L$ the period of the optical lattice. In the present work, we consider
the condensate in a cigar-type trapped potential with the transversal frequency $\omega_{\bot}$ much larger than
the longitudinal one. The two-component wavefunction $\Psi(x)$ is normalized in
units of $\sqrt{n_0}$, such that the normalization condition turns out to be
\begin{align}\label{normal}
\frac{1}{\pi}\int_{-\pi/2}^{\pi/2}dx(|\Psi_1|^2+|\Psi_2|^2)=1.
\end{align}

The dimensionless GP system \eqref{GP} can also be viewed as a Hamiltonian system governed by the grand canonical
Hamiltonian
\begin{align}\label{grand-canonical}
H=&\frac{1}{\pi}\int_{-\pi/2}^{\pi/2}dx\Bigg\{\Psi^{\dag}\Big[\hat{H} _{\rm{soc}}+V_0\sin^2(x)\Big]\Psi\nonumber\\
&+\frac{g_{11}}{2}|\Psi_1|^4+\frac{g_{22}}{2}
|\Psi_2|^4+g_{12}|\Psi_1|^2|\Psi_2|^2-\mu|\Psi|^2\Bigg\},
\end{align}
where $\mu$ is the chemical potential. Among all possible solutions of equation \eqref{GP}, there are states which still have the form of Bloch waves, i.e.,
$\Psi(x,t) =\psi(x)\exp(-i\mu t)=\varphi_k(x)\exp(ikx-i\mu t)$, where $\varphi_k(x)$ is a periodic function of period $\pi$ and $k$ is the Bloch wave number. The solutions $\psi(x)=\left(\psi_1(x),\psi_2(x)\right)^T$ can be found by extremizing the
Hamiltonian \eqref{grand-canonical} and  satisfy the stationary GP equation,
\begin{align}\label{stationaryGP}
\left[\hat{H} _{\rm{soc}}+V_0\sin^2(x)+g(|\psi_1|^2+|\psi_2|^2)\right]\psi=\mu\psi.
\end{align}
For simplicity, in Eq.~\eqref{stationaryGP}, we
have assumed a SU(2)-symmetric spin
interaction with all equal nonlinearities $g_{11}=g_{22}=g_{12}=g$ unless explicitly stated otherwise.
This assumption is reasonable because the regime where
the nonlinear coefficients are fully tunable for SO-coupled BECs can be achieved in the experiment\cite{JRLi}.

\section{Exact nonlinear Bloch waves}
In the previous work\cite{ZChen}, the ground-state phase diagram of the SO-coupled lattice system \eqref{GP} has been numerically studied,
and an unexpected result is that a single momentum phase (a Bloch wave) with zero longitudinal ($\langle\sigma_z\rangle$) and nonzero
transverse ($\langle\sigma_x\rangle$) spin polarization appears at the the Brillouin zone edge, which has no counterpart in the uniform system. Generally,  it is hard to find exact analytical
solutions for such a complex nonlinear system. However, here we will show that for
sufficiently strong SU(2)-symmetric spin
interactions, there exists a family of generic exact
analytical Bloch solutions to the GP equation \eqref{GP} for a condensate
with a wave vector $k=\pm 1$ corresponding to the boundary of
the first Brillouin zone. Using the trial solution method (see details in Appendix), we obtain the exact
Bloch solution of Eq.~\eqref{stationaryGP}  as follows:
\begin{align}\label{Solutfor psi}
  \left(
    \begin{array}{c}
      \psi_1 \\
      \psi_2 \\
    \end{array}
  \right)
  =&C_1\left(
    \begin{array}{c}
      \sin\frac{\theta}{2} \\
    -\cos\frac{\theta}{2}
    \end{array}\right)e^{ix}
    +C_2\left(
    \begin{array}{c}
      \cos\frac{\theta}{2} \\
    -\sin\frac{\theta}{2}
    \end{array}\right)e^{-ix},\nonumber\\
    C_{1,2}=&\frac{\sqrt{g+\frac{V_0}{2\sin\theta}}\pm\sqrt{g-\frac{V_0}{2\sin\theta}}}{2\sqrt{g}},
\end{align}
with
\begin{equation}\label{theta def}
  \sin\theta=\frac{\Omega}{2\sqrt{\frac{\Omega^2}{4}+k_0^2}},~~
  \cos\theta=\frac{k_0}{\sqrt{\frac{\Omega^2}{4}+k_0^2}}.
\end{equation}
In Eq.~\eqref{Solutfor psi}, the plus in the plus/minus ($\pm$) sign  is for $C_1$, and the minus is for $C_2$.
We can easily prove this exact
solution by directly
substituting Eqs.~\eqref{Solutfor psi} and \eqref{theta def} into
the stationary GP equation \eqref{stationaryGP}. Upon this substitution, we have $\mu=1/2+g+V_0/2-\sqrt{\Omega^2/4+k_0^2}$.
This solution is a nonlinear Bloch wave state at the edge of the Brillouin zone, which only exists when the all-equal nonlinear coefficient is above a critical value, i.e., $ \tilde{g} =
2g\sin\theta/V_0\geq 1$. Apparently, the extra nonlinear Bloch wave state has no counterpart in the linear case.

When $k_0=0$ (in the absence of SO coupling), we have $\sin\theta=1$ and the exact solution \eqref{Solutfor psi} can be reduced to the solution
discovered previously in the regular BEC\cite{Bronski1}
\begin{align}\label{solutionwithoutsoc}
\psi_1=-\psi_2=\sqrt{-\frac{V_0}{2g}\sin^2(x)+\frac{B}{2g}}\exp[i\Theta(x)],
\end{align}
where
\begin{align}\label{solutionwithoutsoc2}
\tan[\Theta(x)]&=\sqrt{1-V_0/B}\tan(x),\nonumber\\
B&=g+V_0/2,~~
\end{align}
with the chemical potential $\mu=B+1/2-\Omega/2$.
Note that the notations of Hamiltonian used here differ slightly from the ones in Ref.~\cite{Bronski1}. If we make the following substitutions:
$g_{11}+g_{22}=2g=1,\Omega=0, V_0\rightarrow -V_0$, Eq.~\eqref{solutionwithoutsoc} exactly recovers  the one [Eq.~(10) of Ref.~\cite{Bronski1}, the solution connecting
the famous loop with the rest of the Bloch band] discovered previously in the regular BEC (without SO coupling) lattice systems.

\begin{figure}[htp]
\center
\includegraphics[width=8cm]{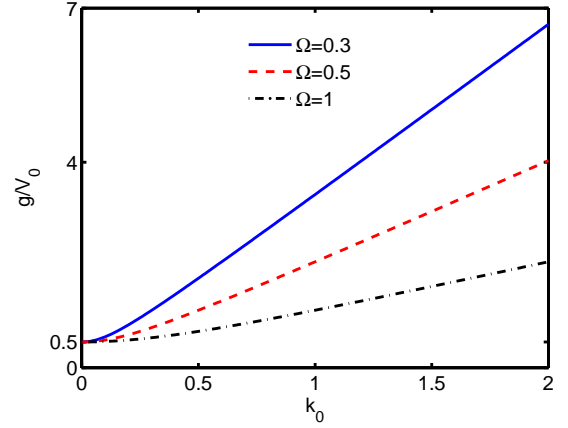}
\caption{(color online) The critical values of $g/V_0$ versus the SO coupling strength $k_0$ for different Rabi frequencies $\Omega=0.3$, $\Omega=0.5$ and $\Omega=1$. The critical lines are given by $ \tilde{g} =
2g\sin\theta/V_0=1$, above which the exact solution \eqref{Solutfor psi} exists for the stationary GP equation \eqref{stationaryGP}. Plotted quantities are in normalized units.} \label{fig1}
\end{figure}

For a more intuitive description of the existence condition of the exact
solution, in Fig.~\ref{fig1} we have plotted the critical lines satisfying $\tilde{g} =1$ for the different parameter
sets. It is clearly seen that the critical values of $g$
monotonically
increase with the SO coupling strength $k_0$ for any fixed Rabi frequency $\Omega$, and increasing Rabi frequency $\Omega$
will lower the critical values of $g$ for the exact solution to exist.  The minimum critical value of $g/V_0=1/2$ exists at $k_0=0$.

\begin{figure}[htp]
\center
\includegraphics[width=8cm]{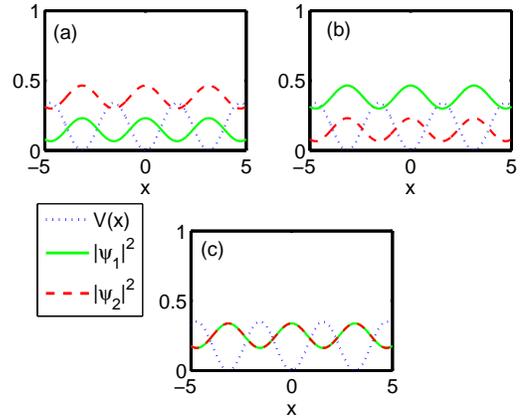}
\caption{(color online) Density profiles along the
$x$ direction and the corresponding potential functions. (a)-(b): $V_0 = 0.34$ satisfying $\tilde{g}>1$, and (c): $V_0 = 0.3511$ at the critical point
$\tilde{g}=1$.  Panel (a) is for the
condensate in the exact Bloch state \eqref{Solutfor psi}, and panel (b) is for the condensate  in the exact Bloch state \eqref{Solutfor psi2}. Solutions \eqref{Solutfor psi} and \eqref{Solutfor psi2}
are twofold degenerate  Bloch states, which coalesce into one single state at the critical point.
All figures show that
the density peaks situate at the valleys of lattice potential wells. The other parameters are $g=0.5,\Omega=0.3,k_0=0.4$. Plotted quantities are in normalized units.} \label{fig2}
\end{figure}

Performing the complex conjugation of
Eq.~\eqref{stationaryGP} and then comparing it with its original equation, we conclude
that for any solution $(\psi_1,\psi_2)^T$ given by Eqs.~\eqref{Solutfor psi} and \eqref{theta def},
$(e^{i\delta}\psi_2^*, e^{i\delta}\psi_1^*)^T$
is also a solution of Eq.~\eqref{stationaryGP} with the same chemical potential for arbitrary global phase $\delta$.
Without loss of generality, we choose $\delta=\pi$ and use notations
of Eqs.~\eqref{Solutfor psi}-\eqref{theta def} to express the other exact solution in the simple form
\begin{align}\label{Solutfor psi2}
  \left(
    \begin{array}{c}
      \psi_1 \\
      \psi_2 \\
    \end{array}
  \right)
  =&C_1\left(
    \begin{array}{c}
      \sin\frac{\theta}{2} \\
    -\cos\frac{\theta}{2}
    \end{array}\right)e^{ix}
    +C_2\left(
    \begin{array}{c}
      \cos\frac{\theta}{2} \\
    -\sin\frac{\theta}{2}
    \end{array}\right)e^{-ix},
    \nonumber\\
    C_{1,2}=&\frac{\sqrt{g+\frac{V_0}{2\sin\theta}}\mp\sqrt{g-\frac{V_0}{2\sin\theta}}}{2\sqrt{g}}.
\end{align}
The solutions \eqref{Solutfor psi} and \eqref{Solutfor psi2} would show explicitly the symmetry by interchange of $C_1$ and $C_2$. That is to say, the solution \eqref{Solutfor psi2} can be obtained by only interchanging the expressions
$C_1$ and $C_2$ of Eq.~\eqref{Solutfor psi}. Solutions \eqref{Solutfor psi} and \eqref{Solutfor psi2} are a pair of degenerate Bloch states to Eq.~\eqref{stationaryGP} for a quasimomentum corresponding to the
boundary of the first Brillouin zone.  From a mathematical point of view, these two extra exact Bloch solutions originate from nonlinear bifurcation.
This leads to a natural speculation that above a critical nonlinearity, swallowtail (or loop) would develop in
the band near the zone boundary for a SO-coupled BEC lattice system, as is the case for the regular BEC in optical lattice.
The looped band structure should be further explored by numerical method.

With the exact solutions, the densities $n_j=|\psi_j|^2$ of each spin component are given by
 \begin{align}
   |\psi_1|^2=&C_1^2\sin^2\frac{\theta}{2}+C_2^2\cos^2\frac{\theta}{2}
   +C_1C_2\sin\theta\cos(2x), \label{psi1 norm} \\
  |\psi_2|^2=&C_1^2\cos^2\frac{\theta}{2}+C_2^2\sin^2\frac{\theta}{2}
   +C_1C_2\sin\theta\cos(2x), \label{psi2 norm}
 \end{align}
and the total density takes the form
\begin{align}\label{psi1 2norm}
   n(x)=&|\psi_1|^2+|\psi_2|^2=1+2C_1C_2\sin\theta\cos(2x),
 \end{align}
where $C_1C_2=V_0/(4g\sin\theta)$.

Equation \eqref{psi1 2norm}
shows that the contrast in $n(x)$ is set by $2C_1C_2\sin\theta=V_0/(2g)$, independent of
the SO coupling strength $k_0$ and Raman frequency $\Omega$. At the critical point $ \tilde{g}=1$, we have $C_1=C_2=1/\sqrt{2}$, and
the two degenerate Bloch states \eqref{Solutfor psi} and \eqref{Solutfor psi2} merge into one single solution with $|\psi_1|^2=|\psi_2|^2$,
which is mathematically
equivalent to the stripe phase (an equal-weight superposition of two plane wave states) existing in the uniform system\cite{YLi}. In Fig.~\ref{fig2} we show the
density profile calculated at $g=0.5,\Omega=0.3,k_0=0.4$.
The two panels (a) and (b) (with $V_0=0.34$) correspond to the exact solutions \eqref{Solutfor psi} and \eqref{Solutfor psi2} respectively, and panel (c)
 corresponds to the exact solution at the critical point $\tilde{g}=1$.
When $\tilde{g}>1$ is set, as shown in Figs.~\ref{fig2} (a) and (b), the spatial distributions for atomic number densities of two components
are  parallel, whose amplitude maxima align to the local minima of optical lattice.
By comparing Figs.~\ref{fig2} (a) with (b), we readily see that the
numbers of atoms in each component for the two exact solutions \eqref{Solutfor psi} and \eqref{Solutfor psi2} are just swapped.
At the critical point $\tilde{g}=1$, the two spin components feature the same spatial distribution [see Fig.~\ref{fig2} (c)].

\begin{figure}[htp]
\center
\includegraphics[width=8cm]{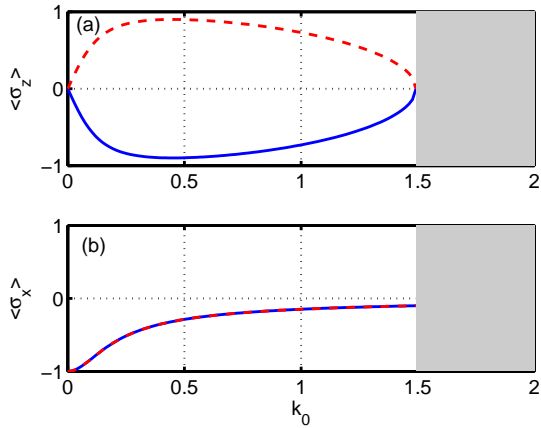}
\caption{(color online) Longitudinal and  transverse
spin polarization $\langle\sigma_z\rangle$ and $\langle\sigma_x\rangle$ as a function
of SO coupling strength $k_0$. The blue lines and red dashed
lines denote spin polarizations of  exact solutions  \eqref{Solutfor psi} and \eqref{Solutfor psi2}  respectively,
and shaded areas correspond to the regions where these exact solutions  \eqref{Solutfor psi} and \eqref{Solutfor psi2} no longer exist.
The other parameters
are $g=0.5, V_0=0.1, \Omega=0.3$. Plotted quantities are in normalized units.} \label{fig3}
\end{figure}

\begin{figure}[htp]
\center
\includegraphics[width=8cm]{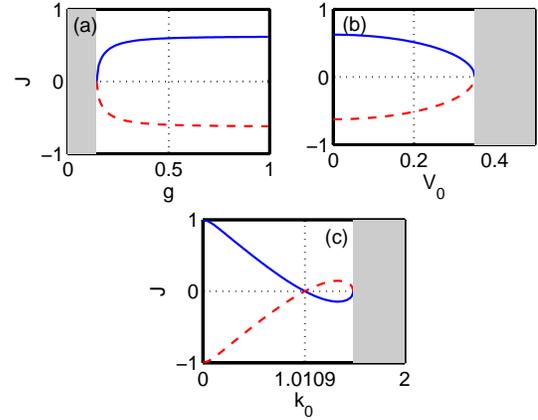}
\caption{(color online) (a) The atomic current density $J$ versus  nonlinearity value $g$. The other parameters
are chosen as $V_0=0.1, k_0=0.4, \Omega=0.3$. (b) The atomic current density $J$ versus lattice depth $V_0$. The other parameters
are chosen as $g=0.5, k_0=0.4, \Omega=0.3$. (c) The atomic current density $J$ versus SO coupling strength $k_0$. The other parameters
are chosen as $g=0.5, V_0=0.1, \Omega=0.3$.
In all plots, the
blue lines and red dashed lines describe the current densities of exact solutions  \eqref{Solutfor psi} and \eqref{Solutfor psi2} respectively,
and shaded areas correspond to the regions where these exact solutions  \eqref{Solutfor psi} and \eqref{Solutfor psi2} no longer exist. Plotted quantities are in normalized units.} \label{fig4}
\end{figure}

To put the exact Bloch states
into perspective, we further calculate the key physical
quantities such as the longitudinal ($\langle\sigma_z\rangle$) and
transverse ($\langle\sigma_x\rangle$) spin polarization of the gas
\begin{align}
  \langle\sigma_z\rangle=&\frac{1}{\pi}\int_{-\pi/2}^{\pi/2}(|\psi_1|^2-|\psi_2|^2)dx
            =\cos\theta(|C_2|^2-|C_1|^2),\label{long gas}\\
\langle\sigma_x\rangle=&\frac{1}{\pi}\int_{-\pi/2}^{\pi/2}(\psi_1^*\psi_2+\psi_1\psi_2^*)dx=-\sin\theta. \label{trans gas}
\end{align}
The longitudinal spin polarizations characterizing
the two exact Bloch states \eqref{Solutfor psi} and \eqref{Solutfor psi2} are then given by the simple expression  $\langle\sigma_z\rangle=\pm\cos\theta\sqrt{1-V_0^2/(4g^2\sin^2\theta)}$ [$-$ for solution \eqref{Solutfor psi} and
$+$ for solution \eqref{Solutfor psi2}], while the transverse
polarizations are given by the same value $\langle\sigma_x\rangle=-\sin\theta$.
This implies that the two exact Bloch states \eqref{Solutfor psi} and \eqref{Solutfor psi2} have opposite longitudinal spin polarizations
and yet identical transverse
polarizations. At the critical point $\tilde{g}=1$, when the condensate is in the single-momentum phase of exact Bloch state with $C_1=C_2=1/\sqrt{2}$  at the Brillouin zone edge, the
longitudinal spin polarization identically vanishes: $\langle\sigma_z\rangle=0$, while $\langle\sigma_x\rangle\neq0$.
This unique single-momentum phase has no analog in homogeneous systems, which seems reminiscent of the Bloch state at the edge of the Brillouin
zone numerically found in the same SO-coupled BEC lattice system\cite{ZChen}.

In Fig.~\ref{fig3}, we plot the  longitudinal and the transverse spin polarizations
$\langle\sigma_z\rangle$ [Fig.~\ref{fig3}(a)] and $\langle\sigma_x\rangle$ [Fig.~\ref{fig3}(b)] as a function of the SO coupling strength $k_0$. As shown in Fig.~\ref{fig3}, when $\tilde{g}>1$ (on the left side of the shaded area), the exact Bloch wave states \eqref{Solutfor psi} and \eqref{Solutfor psi2}  with wave vector $k=1$
describe the single-momentum phase with both nonzero longitudinal and transverse spin polarizations. Here, the red dashed lines correspond to  the spin polarization of the exact Bloch wave state \eqref{Solutfor psi} and the blue solid lines correspond to  the spin polarization of the exact Bloch wave state \eqref{Solutfor psi2}. An exceptional point is that at $k_0=0$ (in the absence of SO coupling), $\langle\sigma_z\rangle$ vanishes, which indicates the longitudinal spin polarization arising from SO coupling.  However, as the SO coupling strength $k_0$ increases to the boundary of the shaded area [that is, the critical point $\tilde{g}=1$ is reached], the system enters a new phase characterized by the merged exact Bloch wave state with  $\langle\sigma_z\rangle=0$ and  $\langle\sigma_x\rangle\neq 0$.

As is well known, all the linear Bloch waves at the zone edge
carry no currents, because the flow $\exp(ix)$ for free particle is stopped completely by Bragg scattering
from the periodic potential. For the regular BEC without SO coupling, when the nonlinearity is strong enough to dominate  the competition
between the  periodic potential and the nonlinear
mean-field interaction, the flow
 can no longer be stopped by Bragg scattering, leading to a current-carrying  Bloch wave at
the edge. Such a superfluidity is related to
the
appearance of a loop (also called swallowtail) structure in the
energy dispersion. And then a question naturally arises: what new effects will the superflow  have in the SO-coupled BEC lattice system?

With the exact solutions at hand, we can directly examine the superfluidity. From the continuity equation,
$(d/dt)n(x)+\nabla\cdot \vec{J}=0$ (in our case, $\vec{J}=J\vec{i}$),
it follows that the atomic current density
of the condensate can be written as
\begin{align}\label{flow dens1}
  J=&\frac{i}{2}\left(\frac{d\psi^\dag}{dx}\psi-\psi^\dag\frac{d\psi}{dx}\right)-k_0\psi^\dag\hat{\sigma}_z\psi\nonumber\\
  =&(|C_1|^2-|C_2|^2)(1-k_0\cos\theta),
\end{align}
where the second part
in  current density is induced by the SO coupling. Note that
Eq.~\eqref{Hsoc} is an effective Hamiltonian that describes SO coupling in the
frame transformed via the local pseudospin rotation.
To look into the current density \eqref{flow dens1} in  laboratory reference frame,
 we apply a unitary transformation to the wave functions  $\phi_1=\psi_1e^{ik_0x}$ and $\phi_2=\psi_2e^{-ik_0x}$, and the original Hamiltonian
in laboratory frame for $\phi=(\phi_1, \phi_2)^T$ is $H_{\rm{lab}}=\frac{\hat{p}_x^2}{2}+\frac{\Omega}{2}(e^{i2k_0x}\hat{\sigma}_++e^{-i2k_0x}\hat{\sigma}_-)+V_0\sin^2(x)+\hat{H} _{\rm{non}}$, where $\hat{\sigma}_{\pm}=\hat{\sigma}_{x}\pm i \hat{\sigma}_{y}$. When moving back to the laboratory
frame, we can employ the conventional current density $J=\frac{i}{2}\left(\frac{d\phi^\dag}{dx}\phi-\phi^\dag\frac{d\phi}{dx}\right)$ to yield
the same formula  \eqref{flow dens1}. The result is obvious because all the physical observable quantities are independent of reference frame.

In Fig.~\ref{fig4}, we plot the atomic current density $J$ versus the system parameters such as the nonlinearity value $g$, lattice depth $V_0$, and SO coupling strength $k_0$.
As shown in Fig.~\ref{fig4}, the two exact solutions  \eqref{Solutfor psi} and \eqref{Solutfor psi2} represent the flow of non-zero speed with opposite signs except at the critical point $\tilde{g}=1$ (at the boundaries of the shaded areas) as  illustrated by the blue lines and red dashed lines respectively.
 These two exact states share the same crystal momentum $k=1$ but have different atomic current velocities, where the solution  \eqref{Solutfor psi}  corresponds to the fluid moving to the right, whereas the other solution  \eqref{Solutfor psi2} corresponds to fluid
moving towards the left.  The current-carrying Bloch wave states at the Brillouin zone edge  are a manifestation of superfluidity.
At the critical point $\tilde{g}=1$ [i.e., at the boundaries of the shaded areas],  the current densities $J$ of these two exact states
identically disappear. This follows because $|C_1|^2=|C_2|^2$ at the critical point.  In the absence of SO coupling ($k_0=0$),  the two exact solutions  \eqref{Solutfor psi} and \eqref{Solutfor psi2} carry  nonzero total density currents, $J=\pm\sqrt{4g^2-V_0^2}/(2g)$.
Another nontrivial finding is that at a particular SO coupling strength $k_0=1.0109$ corresponding to $k_0\cos\theta=1$ [marked by vertical line in Fig.~\ref{fig4}(c)], the two atomic current densities vanish
due to the competition between nonlinearity and SO coupling. This result  occurs uniquely in the SO-coupled BEC lattice system.

Apart from the
continuity equation for total atomic density
current,
there also exists a continuity equation for the spin density and
spin current, which is given by\cite{QFSun}
\begin{align}\label{spin dens1}
\frac{d}{d t} \vec{s}(x, t)=-\nabla \cdot \mathbf{j}_{s}(x, t)+\vec{j}_{\omega}(x, t).
\end{align}
In Eq.~\eqref{spin dens1}, the  spin
density $\vec{s}$ is defined as
$\vec{s}(x, t)=\Psi^{\dagger}\hat{\vec{s}}\Psi$ with $\hat{\vec{s}}=(1/ 2) \hat{\vec{\sigma}}$
and $\hat{\vec{\sigma}}=\hat{\sigma_x}{\vec{i}}+\hat{\sigma_y}{\vec{j}}+\hat{\sigma_z}{\vec{k}}$. The tensor $\mathbf{j}_{s}$ and the vector $\vec{j}_{\omega}$ are named the linear and the angular spin current densities respectively\cite{QFSun}. Here we only focus on the quantity $\mathbf{j}_{s}$
which
describes the translational motion of a spin. As for our model, the linear  spin current densities
take the form
\begin{align}\label{spin dens2}
\mathbf{j}_{s}(x, t)&=\operatorname{Re}\left\{\Psi^{\dagger}\hat{\vec{v}} \hat{\vec{s}}\Psi\right\}=\operatorname{Re}\left\{\Psi^{\dagger}[\hat{p_x}\vec{i}+k_0\hat{\sigma_z}\vec{i}]\hat{\vec{s}}\Psi\right\}.
\end{align}
Thus, the spin current density tensor $\mathbf{j}_{s}^{\nu}$
 ($\nu = x,y,z$ denotes the spin
component) is
\begin{align}\label{spin dens2}
\mathbf{j}_{s}^{\nu}(x, t)=\operatorname{Re}\left\{\Psi^{\dagger}[\hat{p_x}\vec{i}+k_0\hat{\sigma_z}\vec{i}]\hat{s_{\nu}}\Psi\right\}.
\end{align}
By taking
$z$-component of spin as an example, we perform some simple calculations
and derive the analytical spin current density $\mathbf{j}_{s}^{z}$ carried by the exact Bloch states  \eqref{Solutfor psi} and \eqref{Solutfor psi2},
\begin{align}\label{spin dens2}
\mathbf{j}_{s}^{z}(x, t)=-\frac{\cos\theta}{2}\vec{i}+\frac{k_0}{2}\left[1+\frac{V_0}{2g}-\frac{V_0}{g}\sin^2(x)\right]\vec{i}.
\end{align}
From Eq.~\eqref{spin dens2}, we know that these exact Bloch states presented here do carry the spin currents.
According to Eqs.~\eqref{flow dens1} and \eqref{spin dens2},  when $k_0\cos\theta=1$, the total density
current vanishes, while the spin current is nonzero. It is analytically shown that these exact nonlinear Bloch wave states  \eqref{Solutfor psi} and \eqref{Solutfor psi2} carry no total density
current and only pure spin current. This gives an exact demonstration of the pure spin current discussed in the literatures (see, for example, Ref.~\cite{QZhu}).

\begin{figure}[htp]
\center
\includegraphics[width=8cm]{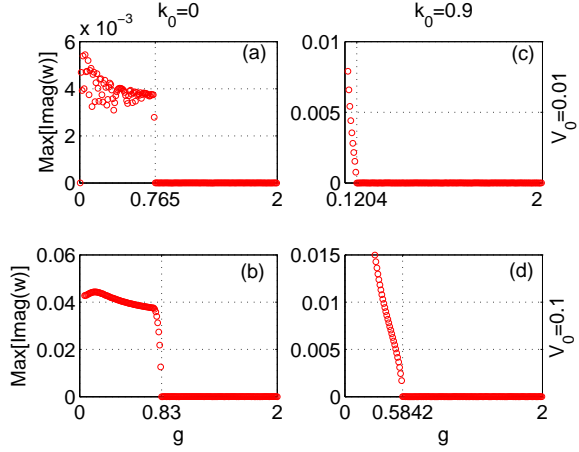}
\caption{(color online) Dynamical stability of the exact Bloch solution \eqref{Solutfor psi} with crystal momentum $k=1$. Maximum imaginary part of $w$ versus nonlinearity strength $g$ for different values of $k_0$ and $V_0$: (a) $k_0=0, V_0=0.01$; (b) $k_0=0, V_0=0.1$; (c) $k_0=0.9, V_0=0.01$; (d) $k_0=0.9, V_0=0.1$. Here $g_{11}=g_{22}=g_{12}=g$ and $\Omega=0.3$. The largest imaginary value
of $w$, $\rm{Max}[\rm{Imag}(w)]$, is used to measure the dynamical stability. $\rm{Max}[\rm{Imag}(w)]=0$ indicates  dynamical stability
and $\rm{Max}[\rm{Imag}(w)]\neq 0$ indicates dynamical instability. Plotted quantities are in normalized units.} \label{fig5}
\end{figure}

\begin{figure}[htp]
\center
\includegraphics[width=8cm]{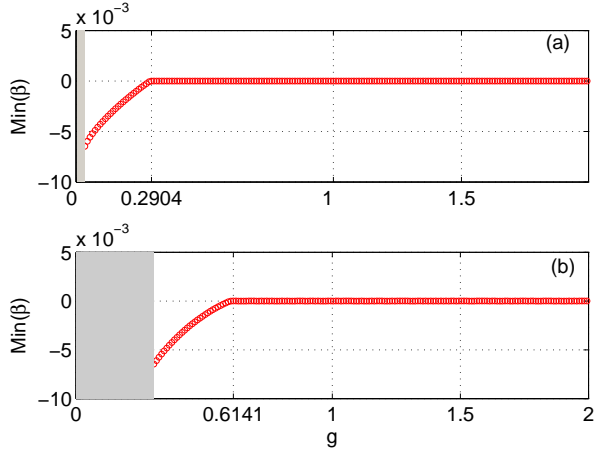}
\caption{(color online) Landau stability of the exact Bloch solution \eqref{Solutfor psi} with crystal momentum $k=1$. (a) Plot of  minimum
(i.e., negative maximum) of $\beta$ versus nonlinearity strength $g$ with the same system parameters as those in \ref{fig5} (c).
(b) Plot of  minimum
(i.e., negative maximum) of $\beta$ versus nonlinearity strength $g$ with the same system parameters as those in \ref{fig5} (d).
  The  minimum value
of $\beta$, $\rm{Min}(\beta)$, is used to measure the Landau stability. $\rm{Min}(\beta)=0$ indicates  Landau stability
and $\rm{Min}(\beta)< 0$ indicates  Landau instability. Shaded areas represent the regions where the exact solution  \eqref{Solutfor psi} no longer exists. Plotted quantities are in normalized units.} \label{fig6}
\end{figure}

So far, our investigations are limited to the case of symmetric spin interaction with all equal nonlinearities $g_{11}=g_{22}=g_{12}$.
We find that for the case of the intercomponent interaction not equal to the
intracomponent interaction, i.e., $g_{11}=g_{22}\neq g_{12}$, when $g_{11}+g_{12}=V_0/\sin\theta$, the GP equation \eqref{GP} also admits exact solution of nonlinear Bloch state at the Brillouin zone edge in the form of $\psi=\frac{1}{\sqrt{2}}e^{ix}(\sin(\theta/2), -\cos(\theta/2))^T+\frac{1}{\sqrt{2}}e^{-ix}(\cos(\theta/2), -\sin(\theta/2))^T$. At present, it remains challenging  to find the more generic exact solutions of nonlinear Bloch states at the Brillouin zone edge  for asymmetric spin interaction, which deserves further investigation in future work.
\section{Stability}
Essential to
the realizability of these above-mentioned exact bloch states  is their stability. For BEC in optical lattices, there are two ways of destroying superfluidity: dynamical instability and Landau instability (also called energetic instability). Once the exact bloch states  of interest are prepared, their stability can be
accurately examined by applying Bogoliubov theory\cite{Wu3}. Assume that the system experiences a small disturbance at a Bloch state:
\begin{align}\label{disturbanceGR}
\Psi= e^{-i\mu t} \left [\psi(x)  + \delta \psi(x,t)   \right ]=e^{-i\mu t+ikx} \left [\varphi_k(x)  + \delta \varphi(x,t)   \right ],
\end{align}
where the disturbance can be written as
\begin{align}\label{disturbance}
 \delta \varphi(x,t) =\begin{pmatrix}
 U_{1}(x) \\
U_{2}(x)
\end{pmatrix} \exp(iqx-iw t) +\begin{pmatrix}
V_{1}^{\ast }(x) \\
V_{2}^{\ast }(x)
\end{pmatrix}\exp(-iqx+iw^{\ast } t),
\end{align}
with $q$ ranging between $-1$ and $1$.
Substituting Eqs.~\eqref{disturbanceGR} and \eqref{disturbance} into the time-dependent Gross-Piteavskii equation \eqref{GP} and
keeping the first-order terms, we obtain the Bogoliubov-de
Gennes (BdG) equation
\begin{align}\label{BdG}
\textbf{M}\left(\begin{array}{c}
U_{1} \\
U_{2} \\
V_{1} \\
V_{2}
\end{array}\right)=w\left(\begin{array}{c}
U_{1} \\
U_{2} \\
V_{1} \\
V_{2}
\end{array}\right)
\end{align}
with the matrix
\begin{widetext}
\begin{align} \label{BdG H}
\textbf{M}=\left(\begin{array}{cccc}
\hat{L}_{+}(k,q) & g_{12} \varphi_{2}^{*} \varphi_{1}+\frac{\Omega}{2} & g_{11} \varphi_{1}^{2} & g_{12} \varphi_{2} \varphi_{1} \\
g_{12} \varphi_{1}^{*} \varphi_{2}+\frac{\Omega}{2} & \hat{L}_{-}(k,q) & g_{12} \varphi_{1} \varphi_{2} & g_{22} \varphi_{2}^{2} \\
-g_{11} \varphi_{1}^{* 2} & -g_{12} \varphi_{2}^{*}\varphi_{1}^{*} & -\hat{L}_{+}^{*}(k,-q) & -g_{12} \varphi_{1}^{*} \varphi_{2}-\frac{\Omega}{2} \\
-g_{12} \varphi_{1}^{*} \varphi_{2}^{*} & -g_{22} \varphi_{2}^{* 2} & -g_{12} \varphi_{2}^{*} \varphi_{1}-\frac{\Omega}{2} & -\hat{L}_{-}^{*}(k,-q)
\end{array}\right),
\end{align}
\end{widetext}
and
\begin{align}\label{BdG l}
\hat{L}_{+}(k,q)=&-\frac{1}{2}\left[\frac{\partial}{\partial x}+i(q+k)\right]^{2} - i k_{0}\left[\frac{\partial}{\partial x}+i(q+k)\right]+V(x)-\mu
\nonumber\\&+2 g_{11}\left|\psi_{1}\right|^{2}+g_{12}\left|\psi_{2}\right|^{2},\nonumber\\
\hat{L}_{-}(k,q)=&-\frac{1}{2}\left[\frac{\partial}{\partial x}+i(q+k)\right]^{2} + i k_{0}\left[\frac{\partial}{\partial x}+i(q+k)\right]+V(x)-\mu
\nonumber\\&+2 g_{22}\left|\psi_{2}\right|^{2}+g_{12}\left|\psi_{1}\right|^{2}.
\end{align}
 Note that the
matrix $M$ is not Hermitian and its eigenvalues $w$ are not necessarily all real. In Eq.~\eqref{BdG}, if any
eigenvalue $w$ has imaginary part, the nonlinear Bloch wave is dynamically
unstable; if otherwise, i.e., eigenvalues $w$ are real for  all $-1\leq q<1$, it is dynamically stable.

The dynamical stabilities of the exact Bloch solution \eqref{Solutfor psi} are numerically computed and shown in Fig.~\ref{fig5} with different values of $k_0$
and $V_0$, where the  dynamical
stability is measured by the largest imaginary value of the eigenvalues $w$.
If the maximum of the imaginary
part of $w$ is
zero, the solution is  dynamically stable; otherwise, certain modes of perturbation grow exponentially with time, and thus the solution is unstable and the superflow breaks down. It is
clear from Fig.~\ref{fig5} that the exact Bloch solution \eqref{Solutfor psi} is dynamically stable
 when
the atomic interaction $g$ is beyond certain critical values (marked by vertical lines).  The critical values of $g$, above which the exact Bloch wave  \eqref{Solutfor psi} is dynamically stable, decrease with increasing the SO coupling from $k_0=0$ (absence of SO coupling) to $k_0=0.9$ as we see in each row.
On the other hand, the critical value of $g$ for
the dynamical stability increases with the lattice potential
strength $V_0$, as can be seen by comparison of the two rows of Fig.~\ref{fig5}.

The onset of Landau
instability can be investigated numerically by solving the
BdG equation $\tau_z\textbf{ M }\textbf{u}=\beta \textbf{u}$ with $\tau_z=\hat{\sigma}_z\otimes I$ and $\textbf{u}=(U_1,U_2,V_1,V_2)^T$.
If any of eigenvalues $\beta$ is negative, elementary excitations associated with the
perturbations will be energetically preferable, causing Landau
instability (superfluidity is lost). In  Fig.~\ref{fig5}, the
minimum (i.e., negative maximum) of $\beta$ is plotted, where nonzero values indicate
the Landau instability and zero values indicate
the Landau stability.  There  also exist  critical values of $g$ beyond which
the Bloch wave \eqref{Solutfor psi} is  Landau stable and it
represents a superflow. For comparison, the system parameters in
Figs.~\ref{fig6} (a) and (b) are set to be the same as those in Figs.~\ref{fig5} (c) and (d) respectively.
By comparison we find that the critical values of $g$ for dynamical stability are smaller than the counterparts  for  Landau stability.
This means that the exact Bloch wave  \eqref{Solutfor psi} with dynamical instability must be Landau instable, whereas the Bloch wave with Landau instability is not necessarily dynamically unstable. These properties are in analogy to the ones of regular BEC lattice system\cite{Wu2}.  The same results of stability analysis hold true for the other degenerate
Bloch wave  \eqref{Solutfor psi2}.

\section{Conclusions}
In conclusion, we have reported simple but generic twofold degenerate exact solutions of current-carrying
nonlinear Bloch states at  the Brillouin zone edge in the
SO-coupled lattice BEC system with a
symmetric spin interaction. These exact solutions occur only above a critical nonlinearity and thus have no linear counterparts.
They are examined in terms of spin polarizations and the condensates superfluid
properties. These two degenerate exact Bloch states are shown to
 share the same crystal momentum but have different atomic flow velocities, one of which corresponds to the fluid moving to the right and the other to fluid
moving to the left. We have also demonstrated that the exact analytical solutions possess some intriguing properties that are absent in the regular BECs without SO coupling loaded into optical lattices. For example, for certain SO coupling strengths, the total density current of the superflow at the Brillouin zone edge is zero, but the spin current is nonzero.

From a
mathematical perspective, these extra  exact nonlinear Bloch solutions emerge due to bifurcation, which makes the nonlinear Bloch band different
from the linear case. It leads to a natural speculation that this SO-coupled BEC lattice system
would develop a looplike structure at the edge  of
the nonlinear Bloch band extensively found in the regular BEC lattice systems. Recently,
the emergence of looplike structure has been reported numerically  in the nonlinear dispersion
relation of the SO-coupled BEC in the uniform space\cite{YongpingZhang2019}. However, our exact solutions provide, for the first time to
our knowledge, an analytical evidence for the looplike Bloch band structure in the SO-coupled BEC lattice system, which will give us an
intuitive insight to help us understand the superfluity and other related properties such as hysteresis, Bloch oscillation and nonlinear
Landau-Zener tunneling. For the purpose of application,  we have analyzed the collective excitations about these exact nonlinear Bloch states
and investigated the superfluidity stability  through dynamical and
Landau stability analysis.

\acknowledgments
The work was supported by the National Natural Science Foundation of China (Grants No.11975110, No.11764022, No.11465009, and No.11947082), the Zhejiang Provincial Natural Science Foundation of China (Grant No. LY21A050002 and No. LZ20A040002), the Scientific and Technological Research Fund of Jiangxi Provincial Education Department (Grants No.GJJ180559, No.GJJ180581, No.GJJ180588, No.GJJ190549, and No.GJJ190577), and Open Research Fund Program of the State Key Laboratory of Low-Dimensional Quantum Physics (Grant No. KF201903). Yunrong Luo was supported by the Scientific Research Fund of Hunan Provincial Education Department
under Grant No. 18C0027 and the National Natural Science Foundation of China under Grants 11747034.

\section*{Appendix}
\label{app}
\setcounter{equation}{0}
\renewcommand{\theequation}{A.\arabic{equation}}
In Appendix, we will give the detailed derivations
of the exact solution to the time-independent GP
equation \eqref{stationaryGP} in the main text.

To proceed, we set the test solution of the time-independent GP
equation \eqref{stationaryGP} as
\begin{equation}\label{testSolutfor psi}
  \left(
    \begin{array}{c}
      \psi_1 \\
      \psi_2 \\
    \end{array}
  \right)
  =C_1\left(
    \begin{array}{c}
      \sin\frac{\theta}{2} \\
    -\cos\frac{\theta}{2}
    \end{array}\right)e^{ix}
    +C_2\left(
    \begin{array}{c}
      \cos\frac{\theta}{2} \\
    -\sin\frac{\theta}{2}
    \end{array}\right)e^{-ix},
\end{equation}
where $C_1, C_2$ and $\theta$ are undetermined parameters.
The superposition
coefficients $C_1=|C_1|e^{i\alpha_1}$ and $C_2=|C_2|e^{i\alpha_2}$ satisfy
 the normalization constraint $|C_1|^2+|C_2|^2=1$ [corresponding to the normalization $(1/\pi)\int_{-\pi/2}^{\pi/2}dx(|\psi_1|^2+|\psi_2|^2)=1$].
It follows from \eqref{testSolutfor psi}  that
 \begin{align}
   |\psi_1|^2=&|C_1|^2\sin^2\frac{\theta}{2}+|C_2|^2\cos^2\frac{\theta}{2}
   +|C_1||C_2|\sin\theta\cos(2x+\beta), \label{testpsi1 norm} \\
  |\psi_2|^2=&|C_1|^2\cos^2\frac{\theta}{2}+|C_2|^2\sin^2\frac{\theta}{2}
   +|C_1||C_2|\sin\theta\cos(2x+\beta), \label{testpsi2 norm}
 \end{align}
where $\beta=\alpha_1-\alpha_2$ is the phase of $C_1C_2^*$. Then we have
 \begin{align}\label{testpsi1 2norm}
   |\psi_1|^2+|\psi_2|^2
   =1+2|C_1||C_2|\sin\theta-4|C_1||C_2|\sin\theta\sin^2\left (x+\frac{\beta}{2}\right ),
 \end{align}
where $|C_1|^2+|C_2|^2=1$ has been used.  It can  be easily verified that under the balanced conditions
$4g|C_1||C_2|\sin\theta=V_0$ and $\beta=0$,
the following  relation is  automatically established
\begin{align}\label{Vcondi}
   g(|\psi_1|^2+|\psi_2|^2)+V_0\sin^2(x)=\frac{V_0}{2}+g.
 \end{align}
The requirements of the two equations $4g|C_1||C_2|\sin\theta=V_0$ and  $|C_1|^2+|C_2|^2=1$ yield
\begin{align}
    |C_1|^2=&\frac{1}{2}\pm\frac{1}{2}\sqrt{1-\frac{V_0^2}{4g^2\sin^2\theta}}, \label{Gen-solu C1} \\
    |C_2|^2=&\frac{1}{2}\mp\frac{1}{2}\sqrt{1-\frac{V_0^2}{4g^2\sin^2\theta}}.\label{Gen-solu C2}
 \end{align}
As long as the modulus of coefficients $C_1$
and $C_2$ is fixed by the above two equations \eqref{Gen-solu C1} and \eqref{Gen-solu C2}, and
the relative phase between the coefficients $C_1$
and $C_2$ is by $\beta=0$,  Eq.~\eqref{stationaryGP} in the main text is reduced to the linear Schr\"{o}dinger equation
\begin{align}\label{linearstationaryGP2}
\left[\hat{H} _{\rm{soc}}+\frac{V_0}{2}+g\right]\psi=\mu\psi.
\end{align}
The parameter $\theta$ in the ansatz \eqref{testSolutfor psi} can be determined by the above
 linear Schr\"{o}dinger equation \eqref{linearstationaryGP2}. If the parameter $\theta$ matches the following relationship
\begin{equation}\label{apptheta def}
  \sin\theta=\frac{\Omega}{2\sqrt{\frac{\Omega^2}{4}+k_0^2}},~~
  \cos\theta=\frac{k_0}{\sqrt{\frac{\Omega^2}{4}+k_0^2}},
\end{equation}
the ansatz \eqref{testSolutfor psi} is just the exact solution of Eq.~\eqref{linearstationaryGP2}.
In fact, under condition \eqref{apptheta def},  the wave functions $e^{ix}(\sin(\theta/2), -\cos(\theta/2))^T$ and  $e^{-ix}(\cos(\theta/2), -\sin(\theta/2))^T$   are two-fold degenerate  eigenstates of the linear Schr\"{o}dinger equation \eqref{linearstationaryGP2}.
Therefore, the linear combination
of two degenerate  eigenstates is also an eigenstate of equation \eqref{linearstationaryGP2} with the same eigenvalue.

Given Eqs.~\eqref{Gen-solu C1}-\eqref{Gen-solu C2} and Eq.~\eqref{apptheta def}, together with $\beta=0$,
Eq.~\eqref{testSolutfor psi} constitutes two degenerate exact Bloch solutions to Eq.~\eqref{stationaryGP} at the
Brillouin zone edge $k=1$. From Eqs.~\eqref{Gen-solu C1}-\eqref{Gen-solu C2},
we can readily observe that the superposition
coefficients $C_1$ and $C_2$ are given by (up to a trivial global  phase $\alpha_1=\alpha_2$)
\begin{align}
   C_1=&|C_1|=\left(\frac{1}{2}\pm\frac{1}{2}\sqrt{1-\frac{V_0^2}{4g^2\sin^2\theta}}\right)^{1/2}\nonumber\\
   =&\frac{\sqrt{g+\frac{V_0}{2\sin\theta}}\pm\sqrt{g-\frac{V_0}{2\sin\theta}}}{2\sqrt{g}},
   \label{Gen-solu C11} \\
   C_2=&|C_2|=\left(\frac{1}{2}\mp\frac{1}{2}\sqrt{1-\frac{V_0^2}{4g^2\sin^2\theta}}\right)^{1/2}\nonumber\\
   =&\frac{\sqrt{g+\frac{V_0}{2\sin\theta}}\mp\sqrt{g-\frac{V_0}{2\sin\theta}}}{2\sqrt{g}},\label{Gen-solu C22}
 \end{align}
which exactly correspond to the two exact solutions \eqref{Solutfor psi} and \eqref{Solutfor psi2} in the main text.

\end{document}